\documentclass[a4paper,10pt]{article}
\usepackage{amsfonts}
\usepackage{amstext}
\usepackage{amssymb,amsmath}
\usepackage{graphicx}
\usepackage{epstopdf}
\usepackage[T1]{fontenc}
\usepackage[utf8]{inputenc}
\usepackage{authblk}

\def\R{\Bbb R}
\def\E{\Bbb E}\def\P{\Bbb P}
\def\0{\bold 0}
\def\1{\bold 1}
\def\cc{\bold c}
\def\d{\bold d}

\def\b{\bold b}
\def\g{\bold g}
\def\m{\bold m}
\def\r{\bold r}

\def\A{\bold A}
\def\B{\bold B}

\def\Dv{\bold D}

\def\alphav{\underline{\alpha}}
\def\betav{\underline{\beta}}

\def\piv{\underline{\pi}}

\def\inter{{\text{int}}\,}
\def\part{\cal P}

\title{Estimating parameters of a multipartite loglinear graph model via the EM algorithm}

\author{Marianna Bolla \thanks{marib@math.bme.hu; also at the Center for 
Telecommunication and Informatics, Debrecen University.}}
\author{Ahmed Elbanna \thanks{ahmed@math.bme.hu; also at the MTA-BME 
Stochastic Research Group.}}

\affil{Institute of Mathematics, Budapest University of Technology and Economics, Hungary}

\begin{document}

\maketitle

\section*{Abstract}

We will amalgamate the Rash model (for rectangular binary tables) and the newly introduced $\alpha$-$\beta$ models (for random undirected graphs) in the framework of a semiparametric probabilistic graph model.
Our purpose is to give a partition of the vertices of an observed graph  so that the generated subgraphs and bipartite graphs obey these models, where their strongly connected parameters give multiscale evaluation of the vertices at the same time. In this way, a heterogeneous version of the stochastic block model is built via mixtures of loglinear models and the parameters are estimated with a special EM iteration. 
In the context of social networks, the clusters can be identified with social groups and the parameters with attitudes of people of one group towards people of the other, which attitudes depend on the cluster memberships. The algorithm is applied to randomly generated and real-word data.

\section {\label{intro}Introduction}

So far many parametric and nonparametric methods have been proposed for community detection in networks. In the nonparametric scenario, hierarchical or spectral methods were applied to maximize the two- or multiway Newman--Girvan modularity~\cite{Clauset,Newman10,Fortunato,Bol11}; more generally, spectral clustering tools, based on Laplacian or modularity spectra, proved to be feasible to find community, anticommunity, or regular structures in networks~\cite{Bol13}. In the parametric setup, certain model parameters are estimated, usually via maximizing the \textit{likelihood function} of the graph, i.e., the joint probability of our observations under the model equations. This so-called ML estimation is a promising method of statistical inference, has solid theoretical foundations~\cite{Rao,McLachlan}, and also supports the common-sense goal of accepting parameter values based on which our sample is the most likely.

In the 2010s, $\alpha$-$\beta$-models~\cite{Chatterjee,Csiszar1} were developed as the unique graph models where the degree sequence is a \textit{sufficient statistic}: given the degree sequence, the distribution of the random graph does not depend on the parameters any more 
(microcanonical distribution over the model graphs). This fact makes it possible to derive the ML estimate of the parameters in a standard way~\cite{Rinaldo}. Indeed, in the context of network data, a lot of information is contained in the degree sequence, though, perhaps in a more sophisticated way. The vertices may have \textit{clusters} (groups or modules) and their memberships
may affect their affinity to make ties. We will find groups of the vertices  such that the within- and between-cluster edge-probabilities admit certain parametric graph models, the parameters of which are highly interlaced. Here the degree sequence
is not a sufficient statistic any more, only if it is restricted to the subgraphs. When making inference, we are partly inspired by the stochastic block model, partly by the Rasch model, the rectangular analogue of the $\alpha$-$\beta$ models.

The first type of block models
is the homogeneous one: the probability to make ties is the same
within the clusters or between the cluster-pairs. Although this
probability depends on the actual cluster memberships, given the
memberships of the vertices, the probability that they are connected
is a given constant (parameter to be estimated).
This stochastic block model, sometimes called generalized random graph
or planted partition model, is thoroughly
discussed in~\cite{Holland1,Rohe,Karrer,Choi,Fishkind}.

Here we propose a heterogeneous block model by
carrying on the Rasch model developed more than 50 years ago
for evaluating psychological tests~\cite{Rasch,Rasch1}.
Given the number of clusters and a classification of the vertices,
we will use the Rasch model for the bipartite subgraphs, whereas the
$\alpha$-$\beta$ models for the subgraphs themselves, and process an iteration
(inner cycle) to find the ML estimate of their parameters.
Then, based on the overall likelihood,
we find a new classification of the vertices via taking conditional expectation
and using the Bayes rule. Eventually, the two steps are alternated, giving the
outer cycle of the iteration.
Our algorithm fits into the framework of the EM algorithm, the convergence of
which is proved in exponential families under very general
conditions~\cite{DLR,McLachlan}.
The method was originally developed for missing data, and the name comes from
the alternating \textit{expectation} (E) and \textit{maximization} (M) steps,
where in the E-step (assignment phase)
we complete the data by substituting for the missing
data via taking conditional expectation, while in the
M-step (estimation phase) we find the usual ML estimate of the
parameters based on the so completed data.
The algorithm naturally extends to situations, when not the data itself is
missing, but it comes from a mixture, and the grouping memberships
are the missing parameters.
This special type of the EM algorithm developed for mixtures is often
called collaborative filtering~\cite{Ungar,HP} or
Gibbs sampling~\cite{Casella}, the roots of which method can be traced back
to~\cite{Metropolis}.
In the context of social networks, the clusters can be identified
with social strata and the parameters with attitudes of people of one group
towards people of the other, which attitude is the same for people in the
second group, but depends on the individual in the first group.
The number of clusters is fixed during the iteration, but an initial number
can be obtained by spectral clustering tools.
Together with the proof of the convergence, the algorithm is applied to
randomly generated and real-word data.

This kind of model building is originated both in the statistics literature,
e.g.,~\cite{Holland,Lauritzen,Daudin} and in the physics literature,
e.g.,~\cite{Newman10,New6,New13}.
In~\cite{New0}, the author already considers mixing according to vertex degree.
In~\cite{Karrer} the authors introduce the degree-corrected variant of the
stochastic block model, but they use Poisson edge-probabilities.
In~\cite{New13} the likelihood, depending on Poisson parameters, is
maximized with the trick that first a likelihood
maximization is
performed, then the problem is traced back to the minimum-cut
objective. This is not the EM algorithm, though the idea of mixed tools
resembles that.

In~\cite{Bickel}, without giving an algorithm, the authors maximize the
so-called likelihood modularity over $k$-partitions of vertices,
for given $k$. This is rather a non-parametric way of model fitting,
since, instead of parameters, they substitute the relative frequency of
the edges for their Bernoulli parameters, and theoretically maximize
their profile likelihood with respect to the memberships of the vertices,
which is considered as unknown parameter. They also prove the
consistency of their estimates.
\cite{Reichardt} considers bipartition and multipartition of dense graphs
with arbitrary degree distribution.
In~\cite{Fishkind}, based on the adjacency matrix as a statistical sample,
the authors estimate
the underlying partition of the vertices, given an upper bound for the
number of blocks, in the stochastic block model. They prove that the
suitably modified spectral partitioning procedure is consistent.
Before fitting a model, its complexity may also be investigated.
In~\cite{Escolano}, the authors give the quantification of the
intrinsic complexity of undirected graphs and networks, via
distinguishing between randomness complexity and statistical complexity.

The paper is organized as follows. In Section~\ref{blocks} we describe the
building blocks of our model. In the context of the
$\alpha$-$\beta$ models we
refer to already proved facts about the existence of the ML estimate and
if exists, we discuss the algorithm proposed by~\cite{Csiszar1} together with
convergence facts; while, in the context of the $\beta$-$\gamma$ model,
we introduce
a novel algorithm and prove the convergence of it. In Section~\ref{EM} we use
both of the above algorithms for the subgraphs and bipartite subgraphs of
our sample graph, and we connect them together
in the framework of the EM algorithm.
In Section~\ref{appl}  the algorithm is applied to randomly generated and
real-word data, while in Section~\ref{conc} conclusions are drawn.

\section {\label{blocks}The building blocks}

Loglinear type models to describe contingency tables were proposed, e.g.,
by~\cite{Holland,Lauritzen} and widely used in statistics. Together with
the Rasch model, they give the foundation of our unweighted graph and bipartite
graph models, the building blocks of our EM iteration.
Note that in~\cite{Holland}, the authors also extend their
model to directed graphs.

\subsection{\label{A}$\alpha$-$\beta$ models for undirected random graphs}

With different parameterization,
\cite{Chatterjee} and \cite{Csiszar1} introduced the following
random graph  model, where the degree sequence is a sufficient statistic.
We have an unweighted, undirected random graph on $n$ vertices without loops,
such that edges between distinct vertices come into existence independently, but
not with the same probability as in the classical Erd\H os--R\'enyi
model~\cite{Erdos}.
This random graph can uniquely be characterized
by its $n\times n$ symmetric adjacency
matrix $\A =(A_{ij})$ which has zero diagonal and the
entries above the
main diagonal are independent Bernoulli random variables whose
parameters $p_{ij} =\P (A_{ij}=1)$ obey the following rule.
Actually,  we formulate this rule for the $\frac{p_{ij}}{1-p_{ij}}$ ratios,
the so-called \textit{odds}:
\begin{equation}\label{amodel}
  \frac{p_{ij}}{1-p_{ij}} = \alpha_i \alpha_j
   \quad (1\le i < j\le n),
\end{equation}
where the parameters $\alpha_1 ,\dots ,\alpha_n$ are positive reals.
This model is called $\alpha$ \textit{model} in \cite{Csiszar1}.
With the parameter transformation $\beta_i =\ln \alpha_i$ $(i=1,\dots n)$,
it is equivalent to the $\beta$ \textit{model} of \cite{Chatterjee}
which applies to the \textit{log-odds}:
\begin{equation}\label{bmodel}
  \ln \frac{p_{ij}}{1-p_{ij}} = \beta_i +\beta_j
   \quad (1\le i < j\le n)
\end{equation}
with real parameters $\beta_1 ,\dots ,\beta_n$.

Conversely, the probabilities $p_{ij}$ and $1-p_{ij}$ can be expressed in terms
of the parameters, like
\begin{equation}\label{abeta}
  p_{ij} =\frac{\alpha_i \alpha_j}{1+\alpha_i \alpha_j}
\quad \textrm{and}\quad
  1-p_{ij} =\frac{1}{1+\alpha_i \alpha_j}
\end{equation}
which formulas will be intensively used in the subsequent calculations.

We are looking for the  ML estimate of the parameter vector
$\alphav =(\alpha_1 ,\dots ,\alpha_n)$ or $\betav =(\beta_1 ,\dots ,\beta_n )$
based on the observed unweighted, undirected graph as a statistical sample.
(It may seem that we have a one-element sample here, however,
there are $n \choose 2$ independent random variables, the adjacencies,
in the background.)

Let $\Dv =(D_1 , \dots ,D_n )$  denote
the degree-vector of the above random graph, where $D_i =\sum_{j=1}^n A_{ij}$
$(i=1,\dots n)$. The random vector $\Dv$, as a function of the sample entries
$A_{ij}$'s, is a \textit{sufficient statistic} for the parameter $\alphav$,
or equivalently, for $\betav$.
Roughly speaking, a sufficient statistic itself contains all the
information -- that can be retrieved from the data -- for the parameter.
More precisely, a statistic is sufficient when
the conditional distribution of the sample, given the statistic, does not
depend on the parameter any more.
By the \textit{Neyman--Fisher factorization theorem}~\cite{Rao},
a statistic is
sufficient if and only if the likelihood function of the sample can be
factorized into two parts: one which does not contain the parameter, and the
other, which includes the parameter, contains the sample entries merely
compressed into this sufficient statistic. 
Consider this factorization of the likelihood function
(joint probability of $A_{ij}$'s) in our case. Because of
the symmetry of $\A$, this is
$$
\begin{aligned}
L_{\alphav } (\A ) &=
 \prod_{i=1}^{n-1} \prod_{j=i+1}^n p_{ij}^{A_{ij}} (1-p_{ij})^{1-A_{ij}} \\
&=\left\{ \prod_{i=1}^n \prod_{j=1}^n p_{ij}^{A_{ij}} (1-p_{ij})^{1-A_{ij}}
    \right\}^{1/2} \\
&=\left\{ \prod_{i=1}^n \prod_{j=1}^n\left(\frac{p_{ij}}{1-p_{ij}}
 \right)^{A_{ij}} \prod_{i=1}^n \prod_{j=1}^n (1-p_{ij} ) \right\}^{1/2} \\
&= \left\{ \prod_{i=1}^n \alpha_i^{\sum_{j=1}^n A_{ij}}
 \prod_{j=1}^n \alpha_j^{\sum_{i=1}^n A_{ij}}
  \prod_{i\ne j} (1-p_{ij}) \right\}^{1/2} \\
&= \left\{ \prod_{i\ne j} \frac1{1+\alpha_i \alpha_j} \right\}^{1/2}
   \left\{ \prod_{i=1}^n \alpha_i^{D_i}
 \prod_{j=1}^n \alpha_j^{D_j} \right\}^{1/2}  \\
 &=\left\{ \prod_{i < j} \frac1{1+\alpha_i \alpha_j} \right\}
  \left\{ \prod_{i=1}^n \alpha_i^{D_i} \right\}
  = C_{\alphav } \times \prod_{i=1}^n \alpha_i^{D_i}
\end{aligned}
$$
where we used~(\ref{abeta}) and the facts that $A_{ij} =A_{ji}$,
$p_{ij} =p_{ji}$ $(i<j)$ and $A_{ii}=0$, $p_{ii} =0$ $(i=1,\dots ,n)$.
Here the partition function
$C_{\alphav } =\prod_{i<j} \frac1{1+\alpha_i \alpha_j}$
only depends on $\alphav$, and the whole likelihood function depends on
the $A_{ij}$'s merely through $D_i$'s.
Therefore, $\Dv$ is a sufficient statistic.
The other factor is constantly
1, indicating that the conditional
joint distribution of the entries -- given $\Dv$ -- is uniform, but we will not
make use of this fact.
Note that in \cite{Karrer}, the authors call the uniform distribution on graphs
with fixed degree sequence \textit{microcanonical}.
In \cite{Chatterjee,Rinaldo} the converse statement is also proved:
the above $\alpha$ model (reparametrized as $\beta$ model)
is the unique one, where the degree sequence is a sufficient statistic.

Let $(a_{ij})$ be the matrix of the sample realizations (the adjacency
entries of the observed graph),
$d_i =\sum_{j=1}^n a_{ij}$ be the actual degree of vertex $i$
$(i=1,\dots ,n)$ and $\d =(d_1 ,\dots ,d_n )$ be the observed degree-vector.
The above factorization also indicates that the joint distribution of the
entries belongs to the exponential family, and hence, with natural
parameterization~\cite{DLR},
the maximum likelihood estimate $\hat{\alphav}$ (or equivalently,
$\hat{\betav}$) is derived
from the fact that, with it, the observed degree $d_i$ equals
the expected one, that is $\E (D_i) = \sum_{i=1}^n p_{ij}$. Therefore,
$\hat{\alphav }$ is the solution of the following
\textit{maximum likelihood equation}:
\begin{equation}\label{maxlik}
 d_i =
 \sum_{j\ne i}^n \frac{{\alpha}_i {\alpha}_j}
 {1+{\alpha}_i {\alpha}_j}
 \quad (i=1,\dots ,n).
\end{equation}
The ML estimate $\hat{\betav}$ is easily
obtained from $\hat{\alphav }$ via taking the logarithms of its coordinates.

Before discussing the solution of the system of equations~(\ref{maxlik}),
let us see,
what conditions a sequence of nonnegative integers should satisfy so that it
could be realized as the degree sequence of  a graph.
The sequence $d_1 ,\dots ,d_n$ of nonnegative integers is called
\textit{graphic} if
there is an unweighted, undirected graph on $n$ vertices such that
its vertex-degrees are the numbers $d_1 ,\dots ,d_n$ in some order.
Without loss of generality, $d_i$'s can be enumerated in non-increasing order.
The Erd\H os--Gallai theorem~\cite{ErdosG} gives the following necessary and
sufficient condition for a sequence to be graphic.
The sequence $d_1 \ge \dots \ge d_n \ge 0$ of integers is graphic
if and only if it satisfies the following two conditions:
$\sum_{i=1}^n d_i$ is even and
\begin{equation}\label{bfelt}
  \sum_{i=1}^k d_i \le k(k-1)+ \sum_{i=k+1}^n \min\{ k,d_i \} ,
          \quad k=1,\dots , n-1.
\end{equation}

Note that for nonnegative (not necessarily integer) real sequences a continuous
analogue of~(\ref{bfelt}) is derived in \cite{Chatterjee}. For given $n$,
the convex hull of all possible graphic degree sequences is a polytope,
to be denoted by ${\cal D}_n$. Its extreme points are the so-called
\textit{threshold graphs}~\cite{Mahadev}. It is interesting that for $n=3$
all undirected graphs are
threshold, since there are 8 possible graphs on 3 nodes, and there are also 8
vertices of  ${\cal D}_3$; the $n=2$ case is also not of much interest,
therefore we will treat the $n >3$ cases only.
The number of vertices of ${\cal D}_n$ superexponentially grows
with $n$~\cite{Stanley},
therefore the problem of characterizing threshold graphs
has a high computational complexity. Its facial and cofacial sets are
fully described in~\cite{Rinaldo}.
Apart from the trivial cases (when there is at least one degree equal to 0 or
$n-1$),
in~\cite{Hammer}, the authors give the following equivalent
characterization of a threshold graph for $n\ge 4$: it has no four different
vertices $a,b,c,d$ such that $a,b$ and $c,d$ are connected by an edge, but
$a,c$ and $b,d$ not, i.e., it has no two disjoint copies of the complete
graph $K_2$.

The authors of \cite{Chatterjee,Csiszar1} prove that
${\cal D}_n$ is the topological closure of the set of expected degree
sequences, and for given $n >3$,
if  $\d \in \inter ({\cal D}_n )$ is an
interior point, then the maximum likelihood equation~(\ref{maxlik}) has a
unique solution. Later, it turned out that the converse is also true:
in \cite{Rinaldo} the authors prove that the ML estimate exists
if and only if the observed degree vector is an inner point of ${\cal D}_n$.
On the contrary,
when the observed degree vector is a boundary point of ${\cal D}_n$,
there is at least one  0 or 1 probability $p_{ij}$
which can be obtained only by a parameter vector such that at least one of
the $\beta_i$'s is not finite.
In this case, the likelihood function cannot be maximized with a finite
parameter set, its supremum  is approached with a parameter vector $\betav$
with at least one coordinate tending to $+\infty$ or $-\infty$.
We also remark that,  for `large' $n$, the condition
$\d \in \inter ({\cal D}_n)$ is strongly related to the $\delta$-tameness
condition of~\cite{Bar2}, or to the fact that $\d$ has a `scaling limit' defined
in~\cite{Chatterjee}, also to the notion of `there are no dominant vertices'
of~\cite{Borgs}.

The authors in~\cite{Csiszar1} recommend the
following algorithm and prove that, provided
$\d \in \inter ({\cal D}_n)$, the iteration of it converges to the
unique solution of the system~(\ref{maxlik}).
To motivate the iteration, we rewrite
(\ref{maxlik}) as
$$
 d_i = \alpha_i \sum_{j\ne i} \frac1{\frac1{\alpha_j} +\alpha_i } \quad
 (i=1,\dots ,n).
$$
Then
starting with  initial parameter values
$\alpha_1^{(0)} ,\dots , \alpha_n^{(0)}$ and using the observed
degree sequence $d_1 ,\dots ,d_n$, which is an inner point
of ${\cal D}_n$, the iteration is as follows:
\begin{equation}\label{iter}
 \alpha_i^{(t)} =\frac{d_i}{\sum_{j\ne i}
  \frac1{\frac1{\alpha_j^{(t-1)}} +\alpha_i^{(t-1)} } } \quad (i=1,\dots ,n)
\end{equation}
for $t=1,2,\dots$, until convergence.

\subsection{\label{B}$\beta$-$\gamma$ model for bipartite graphs}


This bipartite graph model
traces back to Haberman~\cite{Haberman},
Lauritzen~\cite{Lauritzen},
and Rasch~\cite{Rasch,Rasch1} who
applied it for psychological and educational
measurements, later market research. The frequently cited Rasch model involves
categorical data, mainly
binary variables, therefore the underlying random object can be thought of as a
contingency table. According to the Rasch model, the entries of an $m\times n$
binary table $\A$ are independent Bernoulli random variables,
where
for the parameter $p_{ij}$ of the entry $A_{ij}$ the following
holds:
\begin{equation}\label{rmodel}
  \ln \frac{p_{ij}}{1-p_{ij}} = \beta_i -\delta_j
   \quad (i=1,\dots m; \, j=1,\dots ,n)
\end{equation}
with  real parameters $\beta_1 ,\dots ,\beta_m$ and
$\delta_1 ,\dots ,\delta_n$.
As an example,
Rasch in~\cite{Rasch} investigated binary tables where the rows
corresponded to persons and the columns to items of some psychological test,
whereas
the $j$th entry of the $i$th row was 1 if person $i$ answered test item $j$
correctly and 0, otherwise. He also gave a description of the parameters:
$\beta_i$ was the ability of person $i$, while $\delta_j$ the difficulty of
test item $j$. Therefore, in view of the model equation~(\ref{rmodel}),
the more intelligent the person and the less
difficult the test, the larger the success/failure ratio was on a
logarithmic scale.

Given an $m\times n$ random binary table $\A =(A_{ij})$, or
equivalently, a bipartite graph, our model is
\begin{equation}\label{bimodel}
  \ln \frac{p_{ij}}{1-p_{ij}} = \beta_i +\gamma_j
   \quad  (i=1,\dots ,m, \, j=1,\dots ,n)
\end{equation}
with 
real parameters $\beta_1 ,\dots ,\beta_m$ and $\gamma_1 ,
\dots ,\gamma_n$; further, $p_{ij} = \P (A_{ij} =1 )$.

In terms of  the transformed parameters
$b_i =e^{\beta_i }$ and  $g_j =e^{\gamma_j }$,
the model~(\ref{bimodel}) is equivalent to
\begin{equation}\label{bimodelb}
  \frac{p_{ij}}{1-p_{ij}} = b_i g_j
   \quad   (i=1,\dots ,m, \, j=1,\dots ,n)
\end{equation}
where $b_1 ,\dots ,b_m$ and $g_1 , \dots ,g_n$ are positive reals.

Conversely, the probabilities can be expressed in terms of the parameters:
\begin{equation}\label{bimod}
  {p_{ij}} =\frac{b_i g_j}{1+b_i g_j } \quad \textrm{and} \quad
  1-{p_{ij}} =\frac{1}{1+b_i g_j } .
\end{equation}

Observe that if (\ref{bimodel}) holds with the parameters
$\beta_i$'s and $\gamma_j$'s,
then it also holds with the transformed parameters $\beta'_i =\beta_i +c$
$(i=1,\dots ,m)$ and $\gamma'_j=\gamma_j -c$ $(j=1,\dots ,n)$
with some $c\in \R$.
Equivalently, if
(\ref{bimodelb}) holds with the positive parameters $b_i$'s and $g_j$'s,
then it also holds with the transformed parameters
\begin{equation}\label{ekvi}
 b'_i =b_i \kappa \quad \textrm{and} \quad
 g_j' =\frac{g_j}{\kappa}
\end{equation}
with some $\kappa >0$. Therefore, the parameters
$b_i$ and $g_j$ are
arbitrary to within a multiplicative constant.

Here the row-sums $R_i =\sum_{j=1}^n A_{ij} $ and the column-sums
 $C_j =\sum_{i=1}^m A_{ij} $ are the sufficient statistics for the parameters
collected in
$\b =(b_1 ,\dots ,b_m )$ and $\g =(g_1 ,\dots ,g_n)$.
Indeed, the likelihood function is factorized as
$$
\begin{aligned}
L_{\b ,\g } (\A ) &=
 \prod_{i=1}^m \prod_{j=1}^n p_{ij}^{A_{ij}} (1-p_{ij})^{1-A_{ij}} \\
&=\left\{ \prod_{i=1}^m \prod_{j=1}^n\left(\frac{p_{ij}}{1-p_{ij}}
 \right)^{A_{ij}} \right\}
 \prod_{i=1}^m \prod_{j=1}^n (1-p_{ij} )  \\
&= \left\{ \prod_{i=1}^m b_i^{\sum_{j=1}^n A_{ij}} \right\}
 \left\{ \prod_{j=1}^n g_j^{\sum_{i=1}^m A_{ij}} \right\}
  \prod_{i=1}^m \prod_{j=1}^n (1-p_{ij})  \\
&= \left\{ \prod_{i=1}^m \prod_{j=1}^n \frac1{1+b_i g_j} \right\}
\left\{ \prod_{i=1}^m b_i^{R_i} \right\}
 \left\{  \prod_{j=1}^n g_j^{C_j } \right\} .
\end{aligned}
$$
Since the likelihood function depends on $\A$ only through its row- and
column-sums, by the Neyman--Fisher factorization theorem,
$R_1 ,\dots ,R_m, C_1 ,\dots ,C_n$
is a sufficient statistic for the parameters.
The first factor (including the partition function) depends only on the
parameters and the row- and column-sums,
whereas the seemingly not present
factor -- which would depend merely on $\A$ -- is constantly
1, indicating that the conditional
joint distribution of the entries, given the row- and column-sums,
is uniform (microcanonical) in this model.
Note that in \cite{Bar2}, the author characterizes random tables sampled
uniformly from the set of 0-1 matrices with fixed margins.
Given the margins, the contingency tables coming from the above
model are uniformly distributed, and a typical table
of this distribution is produced by the $\beta$-$\gamma$ model with
parameters estimated via the row- and column sums as sufficient statistics.
In this way, here we obtain another view of the typical table of \cite{Bar2}.

Based on an observed binary table $(a_{ij})$, since we are
in exponential family, and $\beta_1 ,\dots ,\beta_m , \gamma_1 ,
\dots ,\gamma_n $ are natural parameters,
the likelihood equation is obtained by making  the expectation of the
sufficient statistic equal to its sample value. Therefore, with the notation
$r_i =\sum_{j=1}^n a_{ij} $ $(i=1,\dots ,m)$ and
 $c_j =\sum_{i=1}^m a_{ij} $ $(j=1,\dots ,n)$,
the following \textit{system of likelihood equations} is yielded:
\begin{equation}\label{like}
\begin{aligned}
 r_i &=\sum_{j=1}^n \frac{b_i g_j}{1+b_i g_j }=
       b_i \sum_{j=1}^n \frac{1}{\frac1{g_j} + b_i }, \quad i=1,\dots m; \\
 c_j &=\sum_{i=1}^m \frac{b_i g_j}{1+b_i g_j }=
       g_j \sum_{i=1}^m \frac{1}{\frac1{b_i} +g_j } , \quad j=1,\dots n.
\end{aligned}
\end{equation}
Note that for any sample realization of $\A$,
\begin{equation}\label{feltetel}
 \sum_{i=1}^m r_i =\sum_{j=1}^n c_j
\end{equation}
holds 
automatically. Therefore, there is a dependence between the equations of
the system (\ref{like}), indicating that the solution is not unique,
in accord with our previous remark about the arbitrary scaling factor
$\kappa >0$ of~(\ref{ekvi}).
We will prove that apart from this scaling, the solution is unique if it
exists at all. For our convenience, let $({\tilde \b},{\tilde \g})$ denote
the equivalence class of the parameter vector $(\b ,\g)$, which consists of
parameter vectors $(\b' ,\g' )$ satisfying (\ref{ekvi})
with some $\kappa >0$.
So that to avoid this indeterminacy, we may impose conditions on the
parameters, for example,
\begin{equation}\label{cond}
 \sum_{i=1}^m \beta_i  +  \sum_{j=1}^n \gamma_j =0.
\end{equation}

Like the graphic sequences, here the following sufficient
conditions can be given for the sequences
$r_1 \ge \dots \ge r_m >0$ and $c_1 \ge \dots \ge c_n >0$ of integers
to be row- and column-sums of an $m\times n$ matrix of 0-1 entries
(see, e.g.,~\cite{Bar3}):
\begin{equation}\label{bin}
\begin{aligned}
 \sum_{i=1}^k r_i &\le \sum_{j=1}^n \min \{ c_j ,k\} , \quad k=1,\dots ,m; \\
 \sum_{j=1}^k c_j &\le \sum_{i=1}^m \min \{ r_i ,k\} , \quad k=1,\dots ,n.
\end{aligned}
\end{equation}
Observe that the $k=1$ cases imply
$r_1 \le n$ and $c_1 \le m$; whereas the $k=m$ and $k=n$ cases together imply
$\sum_{i=1}^m r_i =\sum_{j=1}^n c_j$.
This statement is the counterpart of the Erd\H os-Gallai conditions for
bipartite graphs, where -- due to (\ref{feltetel}) -- the
sum of the degrees is automatically even.
In fact, the conditions in (\ref{bin}) are redundant:
one of the conditions -- either the one for the rows,
 or the one for the columns -- suffices
together with (\ref{feltetel}) and  $c_1 \le m$ or $r_1 \le n$.
The so obtained necessary and sufficient conditions define
\textit{bipartite realizable sequences} with the wording of~\cite{Hammer}.
Already in 1957, the author~\cite{Gale} determined arithmetic conditions for
the construction of a 0-1 matrix having given row- and column-sums. The
construction was given via swaps. More generally,
\cite{Ryser} referred to the transportation problem and the Ford--Fulkerson
max flow--min cut theorem~\cite{Ford}.

The convex hull of the bipartite realizable sequences
$\r =(r_1 ,\dots , r_m )$ and $\cc =(c_1 ,\dots , c_n )$
form a polytope
in $\R^{m+n}$, actually, because of (\ref{feltetel}), in an
$(m+n-1)$-dimensional hyperplane of it.
It is called \textit{polytope of bipartite degree sequences} and denoted by
${\cal P}_{m,n}$ in~\cite{Hammer}.
It is the special case of the transportation polytope describing margins
of contingency tables with nonnegative integer entries.
There is an expanding literature on the number of such matrices,
e.g., \cite{BarHar},
and on the number of 0-1 matrices with
prescribed row and column sums, e.g., \cite{Bar1}.

Analogously to 
the considerations of the $\alpha$-$\beta$ models,
and applying the thoughts of the proofs in
\cite{Chatterjee,Csiszar1,Rinaldo}, ${\cal P}_{m,n}$ is the closure of
the set of the expected row- and column-sum sequences in the above model.
In~\cite{Hammer} it is proved that an $m\times n$ binary table, or equivalently
a bipartite graph on the independent sets of $m$ and $n$ vertices, is
on the boundary of ${\cal P}_{m,n}$ if
it does not contain two vertex-disjoint edges.
In this case, the likelihood function cannot be maximized with a finite
parameter set, its supremum  is approached with a parameter vector with
at least one coordinate $\beta_i$ or $\gamma_j$
tending to $+\infty$ or $-\infty$, or equivalently,
with at least one coordinate $b_i$ or $g_j$ tending to $+\infty$ or $0$.
Based on the proofs of~\cite{Rinaldo}, and stated as
Theorem 6.3  in the supplementary material of~\cite{Rinaldo},
the maximum likelihood
estimate of the parameters of model (\ref{bimodelb}) exists if and only if the
observed
row- and column-sum sequence $(\r ,\cc )\in\textrm{ri}\, ({\cal P}_{m,n})$,
the relative interior of ${\cal P}_{m,n}$, satisfying (\ref{feltetel}).
In this case for the probabilities, calculated by the formula
(\ref{bimod}) through the estimated  positive parameter values
${\hat b}_i$'s and ${\hat g}_j$'s
(solutions of(\ref{like})), $0<p_{ij}<1$ holds $\forall i,j$.

Under these conditions, we define an algorithm that converges to the
unique (up to the above equivalence)
solution of the maximum likelihood equation (\ref{like}).
More precisely, we will prove that if
$(\r ,\cc )\in\textrm{ri}\, ({\cal P}_{m,n})$,
then our algorithm gives a unique equivalence
class of the parameter vectors as the fixed point of the iteration,
which therefore provides
the ML estimate of the parameters.

Starting with positive parameter values
$b_i^{(0)}$ $(i=1,\dots ,m)$ and $g_j^{(0)}$ $(j=1,\dots ,n)$
and using the observed row- and column-sums, the
iteration is as follows:
$$
\begin{aligned}
 I. \quad
 b_i^{(t)}&=\frac{r_i}{\sum_{j=1}^n \frac{1}{\frac1{g_j^{(t-1)}}+b_i^{(t-1)}}},
      \quad i=1,\dots m \\
 II. \quad
  g_j^{(t)}&=\frac{c_j}{\sum_{i=1}^m \frac{1}{\frac1{b_i^{(t)}}+g_j^{(t-1)}}},
      \quad j=1,\dots n
\end{aligned}
$$
for $t=1,2,\dots $, until convergence.

To show the convergence, we rewrite the iteration as the series of
$(\phi ,\psi ):\R^{m+n} \to \R^{m+n}$ maps, where
$\phi =(\phi_1 ,\dots ,\phi_m )$ and
$\psi =(\psi_1 ,\dots ,\psi_n )$, further $\psi$ depends on $\phi$
such that
$$
\begin{aligned}
 b_i^{(t)}&=\phi_i (\b^{(t-1)}, \g^{(t-1)} ),
      \quad i=1,\dots m \\
  g_j^{(t)}&=\psi_j ((\b^{(t)}, \g^{(t-1)} ) \\
          &=\psi_j (\phi (\b^{(t-1)}, \g^{(t-1)} )  , \g^{(t-1)} ) ,
      \quad j=1,\dots n .
\end{aligned}
$$

We define
\begin{equation}
 \rho ((\b ,\g ), (\b' ,\g' )) =\max \left\{
  \max \{ \max_{1\le i \le m} \frac{b_i}{b'_i} ,
       \max_{1\le i \le m} \frac{b'_i}{b_i} \},
  \max \{ \max_{1\le j \le n} \frac{g_j}{g'_j},
       \max_{1\le j \le n} \frac{g'_j}{g_j} \}   \right\} .
\label{hosszu}
\end{equation}
It is easy to see that $\rho \ge 1$ and $\rho =1$ if and only if
$(\b ,\g )= (\b' ,\g' )$; further,  $\log \rho$ is a metric.
We will use the following Lemma of \cite{Csiszar1}:
for any integer $n>1$ and arbitrary positive real numbers $u_1 ,\dots ,u_n$ and
$v_1 ,\dots v_n$ we have
$$
 \frac{u_1 +\dots +u_n}{v_1 +\dots +v_n }\le\max_{1\le i \le n} \frac{u_i}{v_i},
$$
and equality holds if and only if the ratios $\frac{u_i}{v_i}$ have the
same value.

Now we prove that the $(\phi ,\psi )$ map is a weak contraction in the
$\log \rho$ metric.
\begin{itemize}
\item
\textit{Step I.} Applying the Lemma twice (first with $n$, then with two
 terms),
$$
\begin{aligned}
\frac{\phi_i (\b ,\g )}{\phi_i (\b' ,\g' )}
  &= \frac{r_i \left( \sum_{j=1}^n \frac{1}{\frac1{g_j}+b_i } \right)^{-1}}
       {r_i \left( \sum_{j=1}^n \frac{1}{\frac1{g'_j}+b'_i } \right)^{-1}}
  = \frac{\sum_{j=1}^n \frac{1}{\frac1{g'_j}+b'_i } }
          {\sum_{j=1}^n \frac{1}{\frac1{g_j}+b_i } }  \\
  &\le \max_{1\le j\le n} \frac{\frac{1}{\frac1{g'_j}+b'_i } }
          {\frac{1}{\frac1{g_j}+b_i } }
   = \max_{1\le j\le n} \frac{{\frac1{g_j}+b_i } }
          {{\frac1{g'_j}+b'_i } }     \\
 &\le
 \max_{1\le j\le n} \max \left\{ \frac{g'_j}{g_j} , \frac {b_i}{b'_i} \right\}
= \max \left\{ \max_{1\le j\le n}\frac{g'_j}{g_j} , \frac {b_i}{b'_i} \right\}.
\end{aligned}
$$
Likewise,
$$
\frac{\phi_i (\b' ,\g' )}{\phi_i (\b ,\g )}  \le
\max_{1\le j\le n} \max \left\{ \frac{g_j}{g'_j} , \frac {b'_i}{b_i} \right\}
= \max \left\{ \max_{1\le j\le n}\frac{g_j}{g'_j} , \frac {b'_i}{b_i} \right\}.
$$
Assume that $\rho ((\b ,\g ), (\b' ,\g' )) =\kappa$ and $\kappa >1$; otherwise,
when $\kappa =1$,
we already have the fixed point and there is nothing to prove.
In view of the above calculations and Equation~(\ref{hosszu}),
$$
  \rho ((\phi (\b ,\g ), \g ) ,(\phi (\b' ,\g' ) , \g')) \le \kappa ,
$$
and the inequality can be attained with equality only if at least one of the
following holds:
\begin{enumerate}
\item
$$
\begin{aligned}
 &(a) \quad
 \max_i \frac{\phi_i (\b ,\g )}{\phi_i (\b' ,\g' )}=\kappa \quad \textrm{or}\\
 &(b) \quad
\max_i \frac{\phi_i (\b' ,\g' )}{\phi_i (\b ,\g )}=\kappa ;
\end{aligned}
$$
\item
\textit{(a)}
 $\max_j \frac{g_j}{g'_j}=\kappa$  \quad \textrm{or} \quad
\textit{(b)}  $\max_j \frac{g'_j}{g_j}=\kappa$.
\end{enumerate}
1(a) is equivalent to: there is an $i$ such that $\frac{b_i}{b'_i}=\kappa$
and $\frac{g'_j}{g_j}=\kappa$, $\forall j$; whereas, 1(b) is equivalent to:
there is an $i$ such that $\frac{b'_i}{b_i}=\kappa$
and $\frac{g_j}{g'_j}=\kappa$, $\forall j$.
1(a) implies 2(b) and 1(b) implies 2(a).
However, it cannot be that 2(a) or 2(b) hold, but 1(a) and 1(b) do not, since
$\max_i \frac{\phi_i (\b ,\g )}{\phi_i (\b' ,\g' )}=\kappa'$ with
$1<\kappa' < \kappa$ would result in $\frac{g'_j}{g_j}=\kappa'$, $\forall j$
that contradicts to 2(b);
likewise, $\max_i \frac{\phi_i (\b' ,\g' )}{\phi_i (\b ,\g )}=\kappa'$ with
$1<\kappa' < \kappa$ would
 result in $\frac{g_j}{g'_j}=\kappa'$, $\forall j$
that contradicts to 2(a).
Therefore, it suffices to keep condition 1.

\item
\textit{Step II.} Again applying the Lemma twice
(first with $m$, then with two terms),
$$
\begin{aligned}
\frac{\psi_j (\phi (\b ,\g ) ,\g)}{\psi_j (\phi(\b' ,\g' ), \g' )}
  &= \frac{c_j \left( \sum_{i=1}^m \frac{1}{\frac1{\phi_i (\b ,\g )}+g_j }
 \right)^{-1}}
       {c_j \left( \sum_{i=1}^m \frac{1}{\frac1{\phi_i (\b' ,\g' )}+g'_j }
 \right)^{-1}}   \\
  &= \frac{\sum_{i=1}^m \frac{1}{\frac1{\phi_i (\b' ,\g' ) }+g'_j } }
          {\sum_{i=1}^m \frac{1}{\frac1{\phi_i (\b ,\g ) }+g_j } }
  \le \max_{1\le i\le m} \frac{\frac{1}{\frac1{\phi_i (\b' ,\g' ) }+g'_j } }
          {\frac{1}{\frac1{\phi_i (\b ,\g ) }+g_j } }   \\
   &= \max_{1\le i\le m} \frac{{\frac1{\phi_i (\b ,\g ) }+g_j } }
          {{\frac1{\phi_i (\b' ,\g')}+g'_j } }  \\
   &\le
 \max_{1\le i\le m} \max \left\{ \frac{\phi_i (\b' ,\g' ) }{\phi_i (\b ,\g ) }
  , \frac {g_j}{g'_j} \right\} \\
&=\max \left\{ \max_{1\le i\le m} \frac{\phi_i (\b' ,\g' ) }{\phi_i (\b ,\g ) }
  , \frac {g_j}{g'_j} \right\} .
\end{aligned}
$$
Likewise,
$$
\begin{aligned}
\frac{\psi_j (\phi (\b' ,\g' ) ,\g')}{\psi_j (\phi(\b ,\g ), \g )} &\le
 \max_{1\le i\le m} \max \left\{ \frac{\phi_i (\b ,\g ) }{\phi_i (\b' ,\g' ) }
  , \frac {g'_j}{g_j} \right\} \\
&=\max \left\{ \max_{1\le i\le m} \frac{\phi_i (\b ,\g ) }{\phi_i (\b' ,\g' ) }
  , \frac {g'_j}{g_j} \right\} .
\end{aligned}
$$
Therefore, in view of Equation~(\ref{hosszu}),
\begin{equation}\label{dupla}
\begin{aligned}
  &\rho ((\phi (\b ,\g ), \psi (\phi (\b ,\g ),\g ) )  ,
        (\phi (\b' ,\g' ), \psi (\phi (\b' ,\g' ),\g' ) ) )  \\
 &\le \rho ((\phi (\b ,\g ), \g ) ,(\phi (\b' ,\g' ) , \g')) \le \kappa
\end{aligned}
\end{equation}
and both inequalities can be attained with equality only if at least one of the
following holds:
\begin{enumerate}
\item
$$
\begin{aligned}
&(a) \quad
 \max_j \frac{\psi_j (\phi (\b ,\g ) ,\g )}{\psi_j (\phi (\b' ,\g' ) ,\g' )}
 =\kappa \quad \textrm{or} \\
&(b) \quad
 \max_j \frac{\psi_j (\phi (\b',\g') ,\g')}{\psi_j (\phi (\b  ,\g  ) ,\g  )}
 =\kappa ;
\end{aligned}
$$
\item
$$
\begin{aligned}
&(a) \quad
\max_i \frac{\phi_i (\b ,\g )}{\phi_i (\b' ,\g' )}=\kappa \quad \textrm{or}\\
&(b) \quad \max_i \frac{\phi_i (\b' ,\g' )}{\phi_i (\b ,\g )}=\kappa .
\end{aligned}
$$
\end{enumerate}
1(a) is equivalent to: there is a  $j$ such that $\frac{g_j}{g'_j}=\kappa$
and $\frac{\phi_i (\b' ,\g' )}{\phi_i (\b ,\g )} =\kappa$,
$\forall i$; whereas, 1(b) is equivalent to:
there is a  $j$ such that $\frac{g'_j}{g_j}=\kappa$
and $\frac{\phi_i (\b ,\g )}{\phi_i (\b' ,\g' )} =\kappa$, $\forall i$.
Here again, 1(a) implies 2(b) and 1(b) implies 2(a), and
it cannot be that 2(a) or 2(b) hold, but 1(a) and 1(b) do not.
Therefore, it suffices to keep condition 1 again.
But conditions I.1 and II.1 together
imply that either
$\frac{b'_i}{b_i}=\kappa$, $\forall i$
and $\frac{g_j}{g'_j}=\kappa$, $\forall j$; or
$\frac{b_i}{b'_i}=\kappa$, $\forall i$
and $\frac{g'_j}{g_j}=\kappa$, $\forall j$.
In either case this means that $(\b ,\g)$ and $(\b' ,\g' )$
belong to the same equivalence
class, and in two steps, we already obtained a fixed point
with due regard to the equivalence classes.
This fixed point cannot be else but the unique solution of the
system of likelihood equations~(\ref{like}), 
which is guaranteed (up to equivalence) if 
$(\r ,\cc )\in\textrm{ri}\, ({\cal P}_{m,n})$.

Otherwise,
both inequalities in (\ref{dupla}) cannot
hold with equality, but there must be a strict inequality.
Consequently,
$$
\begin{aligned}
  &\rho ((\phi (\b ,\g ), \psi (\phi (\b ,\g ),\g ) )  ,
        (\phi (\b' ,\g' ), \psi (\phi (\b' ,\g' ),\g' ) ) )  \\
 &< \rho ((\b ,\g ), (\b' ,\g' )) ,
\end{aligned}
$$
and hence, $f=(\phi,\psi)$ is a weak contraction.
\end{itemize}

Observe that
$f( (\b^{(t  )} ,\g^{(t  )} ) ) =(\b^{(t+1)} ,\g^{(t+1)} )$, and
under the condition $(\r ,\cc )\in\textrm{ri}\, ({\cal P}_{m,n})$,
the ML estimate $({\hat \b},{\hat \g })$ is a unique fixed point of $f$, 
that is
$f({\hat \b},{\hat \g }) =({\hat \b},{\hat \g })$. Therefore, we have
$$
  \ln \rho ((\b^{(t+1)} ,\g^{(t+1)} ),   ({\hat \b},{\hat \g })  )                 
  <   \ln \rho ((\b^{(t  )} ,\g^{(t  )} ), ({\hat \b},{\hat \g })  ) .
$$
This means that $\ln \rho ((\b^{(t)} ,\g^{(t)} ),   ({\hat \b},{\hat \g })  ) $
is a monotonic decreasing sequence of nonnegative entries, and so it has a 
limit $c \ge 0$. But this implies that 
$\lim_{t\to \infty} \ln \rho ((\b^{(t)} ,\g^{(t)} ),   ({\b}^*,{\g }^*)  ) =0$,
where
$({\b}^*,{\g }^*)$ is in the equivalence class of  $({\hat \b},{\hat \g })$,
with scaling constant $\kappa =e^c$.

On the contrary, when $(\r ,\cc )\notin\textrm{ri}\, ({\cal P}_{m,n})$,
the sequence cannot converge to a fixed point, since then it were the
solution of the maximum likelihood equation~(\ref{like}). But we have
seen, that no finite solution can exist in this case. It means that
at least one coordinate of the sequence
$\{ (\b^{(t)} ,\g^{(t)} ) \}$ tends  to infinity. 
We remark, that even in this case, we obtain convergence in the other
coordinates, which issue will emerge when solving the multipartite
graph model, and further discussed in Section~\ref{conc}.

\section {\label{EM}The multipartite graph model}

In the several clusters case, we are putting the bricks together.
The above discussed $\alpha$-$\beta$ and $\beta$-$\gamma$ models will be
the building blocks of a heterogeneous block model.
Here the degree sequences are not any more sufficient for the whole graph,
only for the building blocks of the subgraphs.

Given $1\le k\le n$,
we are looking for $k$-partition, in other words, clusters $C_1, \dots ,C_k$
of the vertices such that
\begin{itemize}
\item different vertices are independently assigned to a cluster $C_u$  with
probability $\pi_u$ $(u=1,\dots ,k)$, where $\sum_{u=1}^k \pi_u =1$;
\item given the cluster memberships, vertices $i\in C_u$ and $j\in C_v$ are
connected  independently, with probability $p_{ij}$ such that
$$
  \ln \frac{p_{ij}}{1-p_{ij}} = \beta_{iv} +\beta_{ju} ,
$$
for any $1\le u,v\le k$ pair. Equivalently,
$$
 \frac{p_{ij}}{1-p_{ij}}=  b_{ic_j} b_{jc_i} ,
$$
where $c_i$ is the cluster membership of vertex $i$
and $b_{iv} = e^{\beta_{iv}}$.
\end{itemize}

The parameters are collected in the vector $\piv =(\pi_1 ,\dots ,\pi_k)$
and the $n\times k$ matrix $\B$ of $b_{iu}$'s $(i\in C_u ,\, u=1,\dots ,k)$.
The likelihood function is the following mixture:
$$
 \sum_{1\le u,v\le k} \pi_u \pi_v \prod_{i\in C_u ,j\in C_v}
 p_{ij}^{a_{ij}}  (1-p_{ij})^{(1-a_{ij})} .
$$
Here $\A =(a_{ij})$ is the incomplete data specification
as the cluster memberships are missing. Therefore,
it is straightforward to use the
EM algorithm,
proposed by~\cite{DLR}, also discussed in~\cite{Hastie,McLachlan},
for parameter estimation from incomplete data.
This special application for mixtures is sometimes called
\textit{collaborative filtering}, see \cite{HP,Ungar},
which is rather applicable to fuzzy clustering.

First we complete our data matrix
$\A$ with latent membership vectors $\m_1 ,\dots ,\m_n$ of the vertices
that are $k$-dimensional i.i.d. $Multy (1,\piv )$ (multinomially distributed)
random vectors. More precisely,
$\m_i =(m_{i1},\dots ,m_{ik} )$, where $m_{iu} =1$
if $i\in C_u$ and zero otherwise.
Thus, the sum of the coordinates of any $\m_i$ is 1,
and $\P (m_{iu} =1 ) =\pi_u$.

Note that, if the cluster memberships where known, then the complete
likelihood would be
\begin{equation}\label{likmod}
 \prod_{u=1}^k  \prod_{i=1}^n \prod_{v=1}^k
 \prod_{j=1}^n [p_{ij}^{m_{jv} a_{ij}}   \cdot
 (1-p_{ij})^{ m_{jv} (1-a_{ij})} ]^{m_{iu}}
\end{equation}
that is valid  only in case of known cluster memberships.

Starting with initial parameter values $\piv^{(0)}$, $\B^{(0)}$ and
membership vectors $\m_1^{(0)} ,\dots ,\m_n^{(0)}$, the $t$-th
step of the iteration is the following ($t=1,2,\dots $).

\begin{itemize}
\item \textbf{E}-step:
we calculate the conditional expectation of each $\m_i$ conditioned
on the model parameters and on the other cluster assignments obtained in step
$t-1$, and collectively denoted by $M^{(t-1)}$.

The responsibility of vertex $i$ for cluster $u$ in the $t$-th step
is defined as the conditional expectation
$\pi_{iu}^{(t)} =\E (m_{iu} \, | \, M^{(t-1)})$,
and by the Bayes theorem, it is
$$
 \pi_{iu}^{(t)} =
 \frac{\P (M^{(t-1)} |m_{iu}=1 ) \cdot \pi_u^{(t-1)} }
{\sum_{v=1}^k \P (M^{(t-1)} |m_{iv}=1 ) \cdot \pi_v^{(t-1)} }
$$
($u=1,\dots ,k; \, i=1,\dots ,n$).
For each $i$, $\pi_{iu}^{(t)}$ is proportional to the numerator,
therefore the conditional probabilities $\P (M^{(t-1)} |m_{iu}=1 )$
should be calculated for $u=1,\dots ,k$.
But this is just the part of the likelihood~(\ref{likmod})
effecting vertex $i$ under the condition $m_{iu}=1$. Therefore,
$$
\begin{aligned}
 &\P (M^{(t-1)} |m_{iu}=1 )   \\
  &= \prod_{v=1}^k
 \prod_{j\in C_v , \, j\sim i} \frac{b_{iv}^{(t-1)} b_{ju}^{(t-1)} }
                           {1+b_{iv}^{(t-1)} b_{ju}^{(t-1)} }
 \prod_{j\in C_v , \, j\nsim i} \frac{1}
                           {1+b_{iv}^{(t-1)} b_{ju}^{(t-1)} }  \\
  &=  \prod_{v=1}^k
 \left\{ \frac{b_{iv}^{(t-1)} b_{ju}^{(t-1)} }
                           {1+b_{iv}^{(t-1)} b_{ju}^{(t-1)} } \right\}^{e_{vi}}
 \left\{ \frac{1}{1+b_{iv}^{(t-1)} b_{ju}^{(t-1)} }
 \right\}^{|C_v| \cdot (|C_v| -1)/2 - e_{vi}}  ,
\end{aligned}
$$
where $e_{vi}$ is the number of edges within $C_v$ that are connected to $i$.

\item \textbf{M}-step:
We update $\piv^{(t)}$ and $\m^{(t)}$:
$\pi_u^{(t)} := \frac1{n} \sum_{i=1}^n \pi_{iu}^{(t)}$ and
$m_{iu}^{(t)} =1$ if $\pi_{iu}^{(t)} =\max_v \pi_{iv}^{(t)}$ and 0,
otherwise (in case of ambiguity, we select the smallest index for the cluster
membership of vertex $i$). This is an ML estimation (discrete one, in the
latter case, for the cluster membership).
In this way, a new clustering of the vertices is obtained.

Then we estimate the parameters in the actual clustering of the vertices.
In the within-cluster scenario, we
use the parameter estimation of model~(\ref{amodel}), obtaining estimates of
$b_{iu}$'s ($i\in C_u$) in each cluster separately $(u=1,\dots ,k)$;
as for cluster $u$,
$b_{iu}$ corresponds to $\alpha_i$ and the number of vertices is $|C_u |$.
In the between-cluster scenario, we use the bipartite graph
model~(\ref{bimodelb}) in the following way. For $u< v$, edges
connecting vertices of  $C_u$ and $C_v$ form a bipartite graph, based on
which the parameters $b_{iv}$ $(i\in C_u )$ and $b_{ju}$ $(j\in C_v )$
are estimated with the above algorithm; here $b_{iv}$'s correspond to $b_i$'s,
$b_{ju}$'s correspond to $g_j$'s, and the number of rows and columns of the
rectangular array corresponding to this bipartite subgraph of $\A$ is
$|C_u |$ and $|C_v |$, respectively.
With the estimated parameters, collected in the $n\times k$ matrix $\B^{(t)}$,
we go back to the E-step, etc.
\end{itemize}
By the general theory of the EM algorithm,
since we are in exponential family, the iteration will converge.
Note that here the parameter $\beta_{iv}$ with $c_i =u$ embodies the affinity
of vertex $i$ of cluster $C_u$ towards vertices of cluster $C_v$; and likewise,
$\beta_{ju}$ with $c_j =v$ embodies the affinity
of vertex $j$ of cluster $C_v$ towards vertices of cluster $C_u$. By the
model, this affinities are added together on the level of the log-odds.
This so-called $k$-$\beta$ model, introduced in \cite{Csiszar2},
is applicable to social networks, where
attitudes of individuals in the same social group (say, $u$) are the same
toward members of another social group (say, $v$), though, this attitude also
depends on the individual in group $u$. 
The model may also be applied to biological networks, where the clusters
consist,
for example, of different functioning synopses or other units of the brain,
see~\cite{Bazso}.

After normalizing the $\beta_{iv}$ $(i\in C_u)$ and $\beta_{ju}$ $(j\in C_v)$
to meet the requirement of~(\ref{cond}) for any $u \ne v$ pair, the
sum of the parameters will be zero, and the sign and magnitude of them
indicates the affinity of nodes of $C_u$ to make ties with the nodes
of $C_v$, and vice versa:
$$
 \sum_{i\in C_u } \beta_{iv} +\sum_{j\in C_v}\beta_{ju} =0 .
$$
This becomes important when we want to compare the parameters corresponding
to different cluster pairs.
For selecting the initial number of clusters we can use considerations
of~\cite{Yan}, while for the initial clustering, spectral clustering tools
of~\cite{Bol13}.

\section{\label{appl}Applications}

Now we illustrate our algorithm via randomly generated and real-world data.
We remark that while processing the iteration, we sometimes run into
threshold subgraphs or bipartite subgraphs on the boundary of the
polytope of bipartite degree sequences. Even in this case our iteration
converged for most coordinates of the parameter vectors, while some
$b_{iv}$ coordinates tended to $+\infty$ or 0 (numerically, when stopping
the iteration, they took on a very `large' or `small' value).
This means that the affinity of node $i$ towards nodes of the cluster $j$
is infinitely `large' or `small', i.e.,  this node is liable to
always or never make ties with nodes of cluster $j$.

First we generated a random graph on $n=580$ vertices and with $k=3$ underlying
vertex-clusters $C_1 ,C_2 ,C_3$ in the following way. Let
$|C_1 |:=190$, $|C_2 |:=193$, $|C_3 |:=197$. The parameters
$\beta_{i1}$ $(i \in C_1)$, $\beta_{i2}$ $(i \in C_1)$, and
$\beta_{i3}$ $(i \in C_1)$  were chosen independently at uniform from the
intervals $[0,1]$, $[-1,1]$, and $[-1,0.5]$, respectively.
The parameters
$\beta_{i1}$ $(i \in C_2)$, $\beta_{i2}$ $(i \in C_2)$, and
$\beta_{i3}$ $(i \in C_2)$  were chosen independently at uniform from the
intervals $[-0.75,0.5]$, $[-1,0]$, and $[-0.5,1]$, respectively.
The parameters
$\beta_{i1}$ $(i \in C_3)$, $\beta_{i2}$ $(i \in C_3)$, and
$\beta_{i3}$ $(i \in C_3)$  were chosen independently at uniform from the
intervals $[-0.25,0.75]$, $[-0.25,0.25]$, and $[-0.5,0.5]$, respectively.

\begin{figure}
\includegraphics[scale=.7]{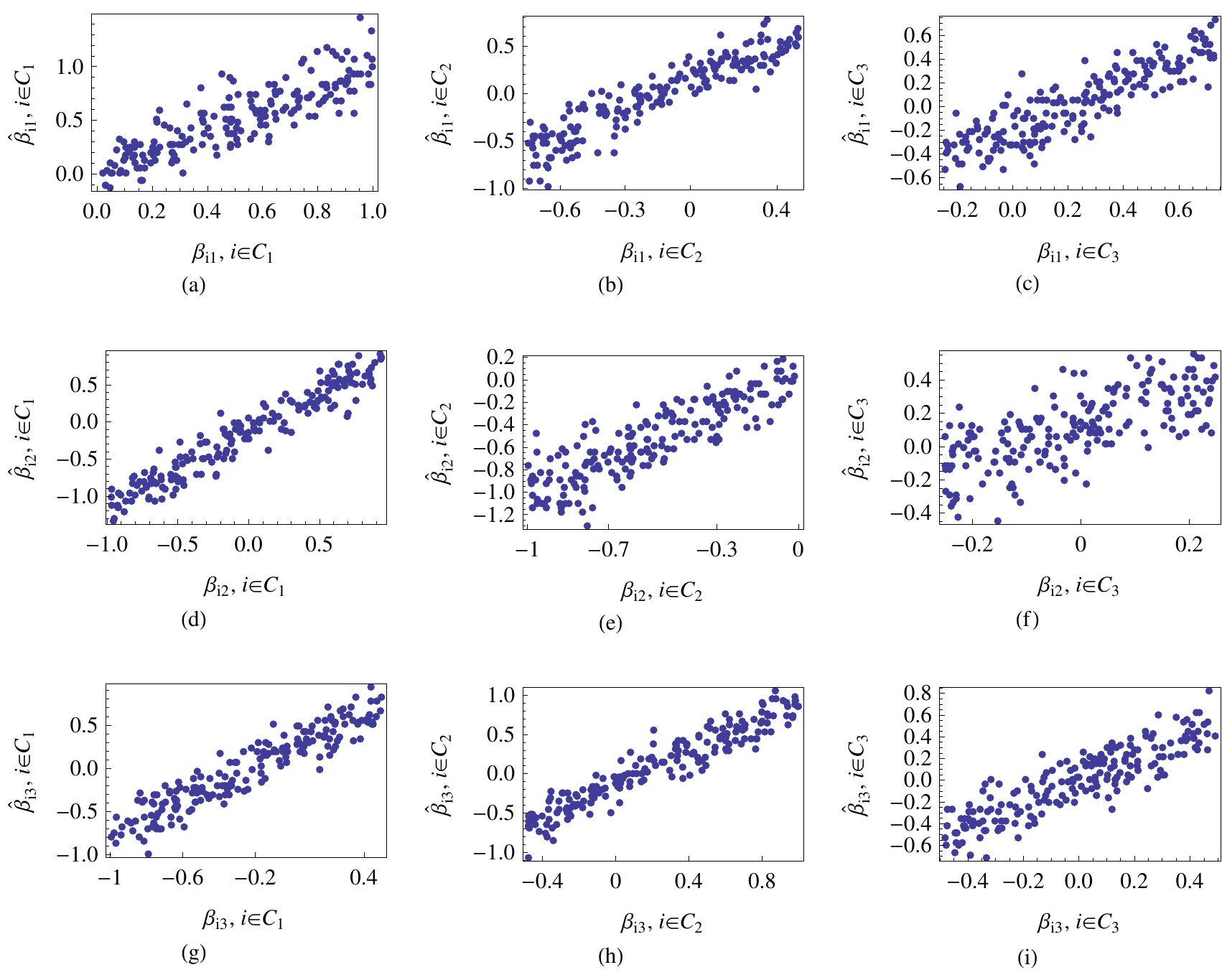}
\caption{
Data were generated based on parameters $\beta_{iv}$'s  chosen uniformly
in different intervals, $k=3$, $|C_1 |=190$, $|C_2 |=193$, $|C_3 |=197$.
The estimated versus the original parameters $\beta_{iv}$'s  are shown for $i \in C_u$ $(u,v=1,\dots ,k)$, where $\beta_{i1} \sim {\cal U}[0,1]$ $(i \in C_1)$, $\beta_{i1} \sim {\cal U}[-0.75,0.5]$ $(i \in C_2)$, $\beta_{i1} \sim {\cal U} [-0.25,0.75]$ $(i \in C_3)$, $\beta_{i2} \sim {\cal U}[-1,1]$ $(i \in C_1)$, $\beta_{i2} \sim {\cal U}[-1,0]$ $(i \in C_2)$, $\beta_{i2} \sim {\cal U}[-0.25,0.25]$ $(i \in C_3)$, $\beta_{i3} \sim {\cal U}[-1,0.5]$ $(i \in C_1)$, $\beta_{i3} \sim {\cal U}[-0.5,1]$ $(i \in C_2)$, and $\beta_{i3} \sim {\cal U}[-0.5,0.5]$ $(i \in C_3)$, respectively.}
\label{simulated}
\end{figure}

Starting with 3 clusters, obtained by spectral clustering tools, and
initial parameter values 
collected in
$\B^{(0)}$ of all 1 entries, after some outer steps, the iteration converged
to $\hat \B =({\hat b}_{iv})$. With ${\hat \beta}_{iv} =\ln {\hat b}_{iv}$,
we plotted the $\beta_{iv} , {\hat \beta}_{iv}$ pairs
for $i\in C_u$, $u,v=1,2,3$.
Fig.~\ref{simulated} shows a good fit of the estimated parameters to the
original ones. Indeed, by the general theory of the ML estimation~\cite{Rao},
for `large' $n$,  the ML estimate should approach
the true parameter, based on which the model was generated.

\begin{figure}
\includegraphics{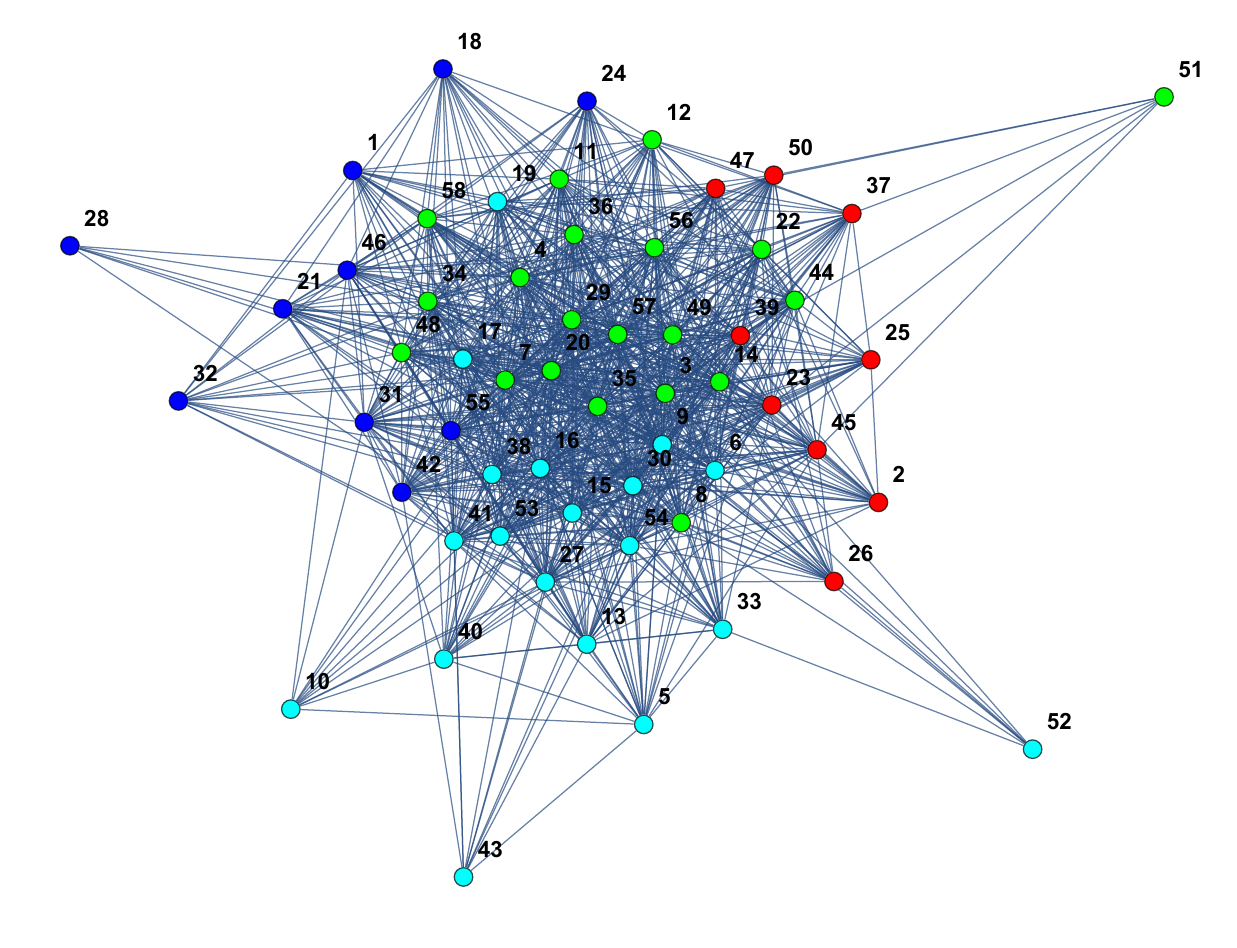}
\caption{
The 4 clusters found by the algorithm in the B\&K fraternity data,
with 10, 9, 20, and 19 students in the clusters, respectively.}
\label{BK}
\end{figure}

Fig.~\ref{BK} shows the resulting clusters obtained by applying our algorithm to the B\&K fraternity data~\cite{BK} with $n=58$ vertices, see also
\textit{http://vlado.fmf.uni-lj.si/pub/networks/data/ucinet/ucidata.htm\#bkfrat}.
 The data, collected by Bernard and Killworth, are behavioral frequency counts, based on communication frequencies between students of a college fraternity in Morgantown, West Virginia. We used the binarized version
of the symmetric edge-weight matrix. When the data were collected, the 58 occupants had been living together for at least three months, but senior students had been living there for up to three years. We used our normalized modularity
based spectral clustering algorithm~\cite{Bol11} to find the starting clusters. In the normalized modularity spectrum we found a gap after the third eigenvalue (in decreasing order of their absolute values), therefore we applied the
algorithm with $k=4$ clusters.  The four  groups are likely to consist to persons living together for about the same time period.

While processing the iteration, occasionally we bumped into the situation when the degree sequence lied on the boundary of the convex polytopes defined in Subsections~\ref{A} and~\ref{B}. Unfortunately, this can occur when our graph is large but not dense enough. In these situations the iteration did not converge for some coordinates $\beta_{iv}$, but they seemed to tend to $+\infty$ or $-\infty$.
Equivalently, the corresponding $\b_{iv}$ for some $i\in C_u$ and $v$ 
tended to $+\infty$ or
0, yielding the situation that member $i\in C_u$ had $+\infty$ or 0 affinity
towards members of $C_v$.
Another way, recommended in~\cite{Chatterjee}, is to add a small amount to each
degree to avoid this situation. However, we did not want to manipulate
the original graph, which was too sparse to produce degree sequences in
the interior of one or more polytopes.

As for the B\&K data, 
Table~\ref{G1}, Table~\ref{G2}, Table~\ref{G3}, and Table~\ref{G4}
show the $b_{iv}$ parameters for $i\in C_1 ,C_1 ,C_3 ,C_4$ and $v=1,2,3,4$.  
The parameters reflecting attitudes of people of the same group towards
each other are usually non-zero finite numbers, whereas there are
many zero or infinite parameters in the intercluster relations.
These may demonstrate that some groups are quite
separated, while some people in some groups show infinite affinity
towards persons of some specific groups.


\begin{table}[h!b!p!]
\begin{center}
\begin{small}
\begin{tabular}{rrrr|rr}
\textbf{Cluster 1} & \textbf{Cluster 2} & \textbf{Cluster 3}  & \textbf{Cluster 4} & \textbf{Label} & \textbf{Degree} \\
\hline
 \text{ }0.30083 & 0 & $\infty$  & 0 & 1 & 23 \\
 \text{ }0.70728 & $\infty$  & $\infty$  & 0 & 18 & 21 \\
 \text{ }1.78066 & $\infty$  & $\infty$  & 0 & 21 & 23 \\
 \text{ }0.12960 & $\infty$  & $\infty$  & 0 & 24 & 26 \\
 \text{ }0 & 0 & $\infty$  & 0 & 28 & 6 \\
 11.14210 & $\infty$  & $\infty$  & $\infty$  & 31 & 39 \\
 \text{ }0.70728 & 0 & $\infty$  & 0 & 32 & 16 \\
 \text{ }0.70728 & $\infty$  & $\infty$  & $\infty$  & 42 & 37 \\
 11.14210 & $\infty$  & $\infty$  & 0 & 46 & 29 \\
 \text{ }4.15633 & $\infty$  & $\infty$  & $\infty$  & 55 & 41 \\
\end{tabular}
\end{small}
\end{center}
\caption{B\&K data: the parameters $b_{iv}$, $i\in C_1$ and $v=1,2,3,4$; 
further, label and degree of vertex $i$, $i \in C_1$.}
\label{G1}
\end{table}

\begin{table}[h!b!p!]
\begin{center}
\begin{small}
\begin{tabular}{rrrr|rr}
\textbf{Cluster 1} & \textbf{Cluster 2} & \textbf{Cluster 3}  & \textbf{Cluster 4} & \textbf{Label} & \textbf{Degree} \\
\hline
 0 & \text{ }0.27583 & $\infty$  & 0.0154392 & 2 & 24 \\
 0 & 15.88360 & $\infty$  & 0.0607916 & 23 & 38 \\
 0 & \text{ }0.27583 & $\infty$  & 0.0413886 & 25 & 25 \\
 0 & \text{ }0.10556 & $\infty$  & 0.0103206 & 26 & 22 \\
 0 & \text{ }2.48471 & $\infty$  & 0.0074528 & 37 & 27 \\
 0 & \text{ }5.86049 & $\infty$  & 0.0875911 & 39 & 38 \\
 0 & 15.88360 & $\infty$  & 0.0592797 & 45 & 34 \\
 0 & \text{ }0.82970 & $\infty$  & 0.0167808 & 47 & 30 \\
 0 & \text{ }0.27583 & $\infty$  & 0.0167808 & 50 & 28 \\
\end{tabular}
\end{small}
\end{center}
\caption{B\&K data: the parameters $b_{iv}$, $i\in C_2$ and $v=1,2,3,4$; 
further, label and degree of vertex $i$, $i \in C_2$.}
\label{G2}
\end{table}

\begin{table}[h!b!p!]
\begin{center}
\begin{small}
\begin{tabular}{rrrr|rr}
\textbf{Cluster 1} & \textbf{Cluster 2} & \textbf{Cluster 3}  & \textbf{Cluster 4} & \textbf{Label} & \textbf{Degree} \\
\hline
 0 & $\infty$  & 1049.5400 & 0 & 3 & 50 \\
 0 & 0 & \text{ }\text{ }\text{ }2.0034 & 0 & 4 & 43 \\
 $\infty$  & 0 & 1049.5400 & 0 & 7 & 52 \\
 0 & 0 & \text{ }\text{ }\text{ }2.0034 & 0 & 8 & 40 \\
 0 & 0 & \text{ }\text{ }\text{ }2.0034 & 0 & 11 & 36 \\
 0 & 0 & \text{ }\text{ }\text{ }1.2482 & 0 & 12 & 29 \\
 0 & 0 & \text{ }\text{ }\text{ }2.0034 & 0 & 14 & 38 \\
 0 & $\infty$  & 1049.5400 & 0 & 20 & 50 \\
 0 & 0 & \text{ }\text{ }\text{ }1.2482 & 0 & 22 & 34 \\
 0 & 0 & \text{ }\text{ }\text{ }8.2779 & 0 & 29 & 46 \\
 $\infty$  & 0 & \text{ }\text{ }\text{ }2.0034 & 0 & 34 & 42 \\
 0 & 0 & \text{ }\text{ }\text{ }8.2779 & 0 & 35 & 47 \\
 0 & 0 & \text{ }\text{ }\text{ }3.5602 & 0 & 36 & 38 \\
 0 & 0 & \text{ }\text{ }\text{ }1.2482 & 0 & 44 & 35 \\
 0 & 0 & \text{ }\text{ }\text{ }0.4075 & 0 & 48 & 31 \\
 0 & 0 & \text{ }\text{ }\text{ }8.2779 & 0 & 49 & 44 \\
 0 & 0 & 0 & 0 & 51 & 6 \\
 0 & 0 & \text{ }\text{ }\text{ }3.5602 & 0 & 56 & 42 \\
 0 & 0 & 1049.5400 & 0 & 57 & 48 \\
 0 & 0 & \text{ }\text{ }\text{ }0.8270 & 0 & 58 & 34 \\
 \end{tabular}
\end{small}
\end{center}
\caption{B\&K data: the parameters $b_{iv}$, $i\in C_3$ and $v=1,2,3,4$; 
further, label and degree of vertex $i$, $i \in C_3$.}
\label{G3}
\end{table}

\begin{table}[h!b!p!]
\begin{center}
\begin{small}
\begin{tabular}{rrrr|rr}
\textbf{Cluster 1} & \textbf{Cluster 2} & \textbf{Cluster 3}  & \textbf{Cluster 4} & \textbf{Label} & \textbf{Degree} \\
\hline
 0 & 10.175 & $\infty$  & 2.73775 & 5 & 24 \\
 141319.0 & 282.644 & $\infty$  & 7.91424 & 6 & 47 \\
 300234.0 & 282.644 & $\infty$  & 4.50778 & 9 & 46 \\
 0 & 4.060 & $\infty$  & 0.24014 & 10 & 12 \\
 52161.5 & 35.602 & $\infty$  & 4.50778 & 13 & 31 \\
 300234.0 & 63.866 & $\infty$  & 2.73775 & 15 & 41 \\
 300234.0 & 121.705 & $\infty$  & 7.91424 & 16 & 44 \\
 300234.0 & 35.602 & $\infty$  & 1.74052 & 17 & 39 \\
 52161.5 & 35.602 & $\infty$  & 0.42163 & 19 & 30 \\
 141319.0 & 63.866 & $\infty$  & 4.50778 & 27 & 42 \\
 141319.0 & 2027.000 & $\infty$  & 2.73775 & 30 & 42 \\
 0 & 4.060 & $\infty$  & 2.73775 & 33 & 27 \\
 141319.0 & 282.644 & $\infty$  & 2.73775 & 38 & 38 \\
 0 & 19.762 & $\infty$  & 0.79274 & 40 & 21 \\
 300234.0 & 35.602 & $\infty$  & 4.50778 & 41 & 40 \\
 0 & 0 & $\infty$  & 0.24014 & 43 & 10 \\
 0 & 4.060 & $\infty$  & 0.06535 & 52 & 7 \\
 52161.5 & 35.602 & $\infty$  & 4.50778 & 53 & 37 \\
 141319.0 & 2027.630 & $\infty$  & 4.50778 & 54 & 44 \\
\end{tabular}
\end{small}
\end{center}
\caption{B\&K data: the parameters $b_{iv}$, $i\in C_4$ and $v=1,2,3,4$; 
further, label and degree of  vertex $i$, $i \in C_4$.}
\label{G4}
\end{table}

\begin{figure}
\includegraphics[width=12.3cm]{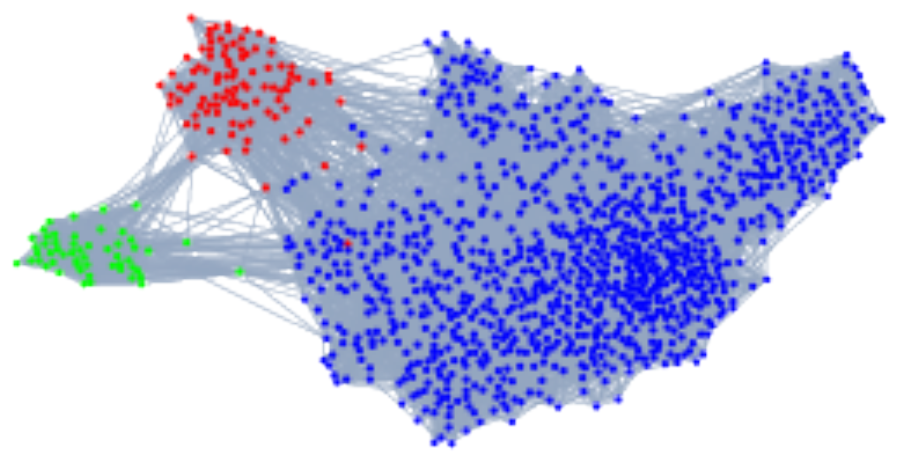}
\caption{
The 3 clusters found by the algorithm in the network of 
the Last.fm users with 1012, 97, and 53 users in the clusters, respectively.}
\label{music}
\end{figure}

We also used the network based on the friendships between the users
of of the Last.fm music recommendation system~\cite{Benczur}.
Last.fm is an online service in music based social networking.
Each user may have friends inside the Last.fm social network,
and so, they form a timestamped undirected graph. In 2012, there were
71,000 users and 285,241 edges between them.
Actually, we only used the
15-core of this graph. 
With spectral clustering tools we found three underlying
clusters, see Fig.~\ref{music}. 

Fragments of the estimated parameter values  
are shown in Table~\ref{M1}, Table~\ref{M2}, and Table~\ref{M3}.
Here there are some zero affinities, but there are no infinite affinities
at all.

\begin{table}[h!b!p!]
\begin{center}
\begin{small}
\begin{tabular}{rrr|rr}
\textbf{Cluster 1} & \textbf{Cluster 2} & \textbf{Cluster 3}  & \textbf{Label} & \textbf{Degree} \\
\hline
 3.099330 & 0.132913 & 0  & 1 & 329 \\
 0.267639 & 0.036161 & 0  & 2 & 45 \\
 0.684764 & 0 & 0 &  3 & 101 \\
 1.174000 & 0 & 0 &  4 & 159 \\
 0.623173 & 0.014063 & 0.038484 &  5 & 95 \\
 0.661509 & 0 & 0 &  6 & 98 \\
 0.886161 & 0 & 0 &  7 & 126 \\
 0.700374 & 0 & 0 &  8 & 103 \\
 0.577860 & 0.013936 & 0 &  9 & 88 \\
 0.446335 & 0 & 0 &  10 & 69 \\
\end{tabular}
\end{small}
\end{center}
\caption{Last.fm data: the parameters $b_{iv}$, $i\in C_1$  and $v=1,2,3$; 
further, label and degree of  vertex $i$, $i \in C_1$ (first 10 users only).}
\label{M1}
\end{table}

\begin{table}[h!b!p!]
\begin{center}
\begin{small}
\begin{tabular}{rrr|rr}
\textbf{Cluster 1} & \textbf{Cluster 2} & \textbf{Cluster 3}  & \textbf{Label} & \textbf{Degree} \\
\hline
 \text{ }1.42735 & 3.398110 & \text{ }\text{ }\text{ }0  & 41 & 68 \\
 \text{ }0.21067 & 0.380953 & \text{ }\text{ }\text{ }0  & 226 & 20 \\
 \text{ }1.16185 & 0.302281 & \text{ }\text{ }\text{ }0  & 263 & 21 \\
 \text{ }0.21067 & 0.566238 & \text{ }\text{ }\text{ }0  & 286 & 26 \\
 \text{ }1.16185 & 0.188533 & \text{ }\text{ }\text{ }0  & 296 & 16 \\
 72.75100 & 0.468577 & \text{ }\text{ }\text{ }5.3444  & 339 & 108 \\
 \text{ }1.16185 & 1.446370 & \text{ }\text{ }\text{ }0.7962  & 340 & 50 \\
 \text{ }0.90754 & 2.845830 & \text{ }\text{ }\text{ }2.3443  & 342 & 65 \\
 \text{ }0.66432 & 1.139340 & \text{ }\text{ }\text{ }0.2664  & 345 & 42 \\
 \text{ }0.66432 & 0.932667 & 372.2770  & 351 & 50 \\
\end{tabular}
\end{small}
\end{center}
\caption{Last.fm data: the parameters $b_{iv}$, $i\in C_2$  and $v=1,2,3$; 
further, label and degree of  vertex $i$, $i \in C_2$ (first 10 users only).}
\label{M2}
\end{table}

\begin{table}[h!b!p!]
\begin{center}
\begin{small}
\begin{tabular}{rrr|rr}
\textbf{Cluster 1} & \textbf{Cluster 2} & \textbf{Cluster 3}  & \textbf{Label} & \textbf{Degree} \\
\hline
 \text{ }\text{ }2.18172 & 0.288845 & 2.370600 &  352 & 43 \\
 \text{ }\text{ }3.60021 & 0.288845 & 2.370600 &  370 & 44 \\
 \text{ }\text{ }0.18727 & 4.071150 & 0.868079 &  379 & 37 \\
 \text{ }\text{ }0.18727 & 0.288845 & 3.359140 &  381 & 44 \\
 \text{ }\text{ }0 & 0.116067 & 0.589030 &  421 & 25 \\
 20.69780 & 0.014618 & 0.589030 &  597 & 34 \\
 \text{ }0.18727 & 0 & 0.482873 &  917 & 22 \\
 0 & 0.538530 & 1.914140 &  942 & 39 \\
 \text{ }0.18727 & 0 & 0.788315  & 953 & 27 \\
 0 & 0 & 1.052870 &  954 & 29 \\
\end{tabular}
\end{small}
\end{center}
\caption{Last.fm data: the parameters $b_{iv}$, $i\in C_3$  and $v=1,2,3$; 
further, label and degree of  vertex $i$, $i \in C_3$ (first 10 users only).}
\label{M3}
\end{table}

\section{\label{conc}Conclusions}

Our model is the heterogeneous version of the
stochastic block model, where the subgraphs and bipartite subgraphs obey
parametric graph models, within which the connections are mainly
determined by the degrees. The EM type algorithm introduced here finds
the blocks and estimates the parameters at the same time.

When investigating controllability of large networks, the authors
of~\cite{Barabasi} observe and prove that a system's controllability
is to a great extent encoded by the underlying network's degree distribution.
In our model, this is true only for the building blocks. Possibly,
the blocks could be controlled separately, based on the degree sequences
of the subgraphs.

Our model is applicable to large inhomogeneous networks, and above finding
clusters of the vertices, it also assigns multiscale parameters to them.
In social networks, these parameters
can be associated with attitudes of persons of one group towards those in
the same or another group. The attitudes are, in fact, affinities to make ties.

\section*{Acknowledgments}

The authors thank G\'abor Tusn\'ady and R\'obert P\'alovics
for fruitful discussions and making the music recommendation data available;
further, Despina Stasi for suggesting us the fraternity data to be processed.
Marianna Bolla's research was supported by the
T\'AMOP-4.2.2.C-11/1/KONV-2012-0001 project and Ahmed Elbanna's research
was partly done under the auspices of the MTA-BME Stochastic Research Group.


\end{document}